\documentclass[aps,pre,floats,floatfix,twocolumn,superscriptaddress]{revtex4}
\usepackage{latexsym,amsmath}
\usepackage{amsbsy,amsfonts}
\usepackage[pdftex]{graphicx}

\usepackage{amssymb}
\usepackage{epstopdf}
\usepackage[usenames,dvipsnames]{xcolor}
\usepackage{soul}
\usepackage[utf8]{inputenc}

\begin{document}

\title{The hidden hyperbolic geometry of international trade:\\ World Trade Atlas 1870-2013}

\author{Guillermo Garc\'ia-P\'erez}
\affiliation{Departament de F{\'\i}sica de la Mat\`{e}ria Condensada, Universitat de Barcelona, Mart\'{\i} i Franqu\`es 1, E-08028 Barcelona, Spain}

\author{Mari\'an Bogu\~{n}\'a}
\affiliation{Departament de F{\'\i}sica de la Mat\`{e}ria Condensada, Universitat de Barcelona, Mart\'{\i} i Franqu\`es 1, E-08028 Barcelona, Spain}

\author{Antoine Allard}
\affiliation{Departament de F{\'\i}sica de la Mat\`{e}ria Condensada, Universitat de Barcelona, Mart\'{\i} i Franqu\`es 1, E-08028 Barcelona, Spain}

\author{M. \'Angeles Serrano}
\affiliation{Departament de F{\'\i}sica de la Mat\`{e}ria Condensada, Universitat de Barcelona, Mart\'{\i} i Franqu\`es 1, E-08028 Barcelona, Spain}
\affiliation{Instituci\'o Catalana de Recerca i Estudis Avan\c{c}ats (ICREA), Passeig Llu\'is Companys 23, E-08010 Barcelona, Spain}

\date{\today}

\begin{abstract}
Here, we present the World Trade Atlas 1870-2013, a collection of annual world trade maps in which distance combines economic size and the different dimensions that affect international trade beyond mere geography. Trade distances, which are based on a gravity model predicting the existence of significant trade channels, are such that the closer countries are in trade space, the greater their chance of becoming connected. The atlas provides us with information regarding the long-term evolution of the international trade system and demonstrates that, in terms of trade, the world is not flat but hyperbolic, as a reflection of its complex architecture. The departure from flatness has been increasing since World War I, meaning that differences in trade distances are growing and trade networks are becoming more hierarchical. Smaller-scale economies are moving away from other countries except for the largest economies; meanwhile those large economies are increasing their chances of becoming connected worldwide. At the same time, Preferential Trade Agreements do not fit in perfectly with natural communities within the trade space and have not necessarily reduced internal trade barriers. We discuss an interpretation in terms of globalization, hierarchization, and localization; three simultaneous forces that shape the international trade system.
\end{abstract}

\maketitle

When it comes to international trade, the evidence suggests that we are far from a distance-free world. Distance still matters~\cite{engel1996} and in many dimensions: cultural, administrative or political, economic, and geographic. This is widely supported by different evidence concerning the magnitude of bilateral trade flows. The gravity model of trade~\cite{tinbergen1962,Bergstrand:1985mw}, in analogy to Newton's law of gravitation, accurately predicts that the volume of trade exchanged between two countries increases with their economic sizes and decreases with their geographical separation. The precision of that model improves when it is supplemented with other factors, such as colony--colonizer relationships, a shared common language, or the effects of political borders and a common currency~\cite{Frankel01052002,Anderson:2004,Bergrstrand:2011}. Despite the success of the gravity model at replicating trade volumes, it performs very poorly at predicting the existence of a trade connection between a given pair of countries~\cite{Duenas:2013}; an obvious limitation that prevents it from explaining the striking regularities observed in the complex architecture of the world trade web~\cite{Serrano:2003kw,Garlaschelli:2004dv,Serrano:2007dg,Serrano:2007nf,serrano:2010}. One of the reasons for this flaw is that the gravity model focuses on detached bilateral relationships and so overlooks multilateral trade resistance and other network effects~\cite{Schweitzer:2009jk}. 

Another drawback of the classical gravity model is that geography is not the only factor that defines distance in international trade.  Here, we use a systems approach based on network science methodologies~\cite{NEWMAN2010,Cohen10_ComplexNetworks} to propose a gravity model for the existence of significant trade channels between pairs of countries in the world. The gravity model is based on economic sizes and on an effective  distance which incorporates different dimensions that affect international trade, not only geography, implicitly encoded on the complex patterns of trade interactions. Our gravity model is based on the connectivity law proposed for complex networks with underlying metric spaces~\cite{Serrano:2008hb,Serrano:2011kq} and it can be represented in a pure geometric approach using a hyperbolic space, which has been conjectured as the natural geometry underlying complex networks~\cite{Boguna:2009uz,KrPa10,Papadopoulos:2012uq,Caldarelli2015pre}. In the hyperbolic trade space, distance combines economic size and effective distance into a sole distance metric, such that the closer countries are in hyperbolic trade space, the greater their chance of becoming connected. We estimate this  trade distance from  empirical data using  adapted statistical inference techniques~\cite{BoPa10,Serrano:2012we}    which allow us to represent international trade through World Trade Maps (WTMs). These define a coordinate system in which countries are located in relative positions according to the aggregate trade barriers between them. The maps are annual and cover a time span of fourteen decades. The collection as a whole, referred to as the World Trade Atlas 1870-2013, is presented via spatial projections~\cite{:vd}, Table S5, and trade distance matrices, Table S6. Beyond the obvious advantages of visualization, the World Trade Atlas 1870-2013 significantly increases our understanding of the long-term evolution of the international trade system and helps us to address a number of important and challenging questions. In particular: How far, in terms of trade, have countries traveled in recent history? What role does each country play in the maps and how have those roles evolved over time? Are Preferential Trade Agreements (PTAs) consistent with natural communities as measured by trade distances? Has the formation of PTAs led to lesser or greater barriers to trade within blocs? Is trade distance becoming increasingly irrelevant? 

The answers to these questions can be summarized by asserting that, in terms of trade, the world is not flat; it is hyperbolic. Differences in trade distances are growing and becoming more heterogeneous and hierarchical; at the same time as they define natural trade communities---not fully consistent with PTAs. Countries are becoming more interconnected and clustered into hierarchical trade blocs than ever before.

%
%
%
%
%
% =================================================================================================
\section*{Mapping the international trade system}
Network representations of world trade~\cite{Serrano:2003kw,Garlaschelli:2004dv,Serrano:2007dg,Serrano:2007nf,serrano:2010} offer a perspective that goes beyond bilateral analysis and allows us to uncover stable large-scale patterns such as the small-world property, heterogeneous distributions of the number of trade partners (degree), and high levels of transitive relationships (clustering). Our main hypothesis is that this architecture is a reflection of the distances between countries in an underlying trade space. We have reconstructed international trade networks using historical aggregate import/export data from two consistent and consecutive databases. For the period 1870--1996, we used data from~\cite{cow,Barbieri:2009}. We then compiled Database S1 which covers the period from 1997 to 2013~\cite{:sm}. In the reconstructed undirected networks, links represent bilateral trade relationships, and the weight of the link corresponds to the value of goods exchanged in a given year, in current US millions of dollars. World war periods, 1914--1919 and 1939--1947, were avoided due to the lack of reported information.
%
%
%
% =================================================================================================
\subsection{Backbones of significant trade channels}

The evolution of trade networks shows trends that are consistent with globalization~\cite{Krugman:1995fi,Baldwin:1999hi,Chase-Dunn:2000sw}, understood as increases in the density of connections, in the total amount traded, and in the relative average geographic length of trade channels (see the Supplementary Information, Fig.~S2~\cite{:sm}). This augmented entanglement obscures patterns and regularities within the networks that we will need to exploit to infer trade distances. In parallel, world trade networks display, especially after World War I (WWI), a strong heterogeneity in the global distributions of: the number of trading partners, total trade per country, and bilateral flows. An increasing heterogeneity is also present at the local level in the distribution of flows between the neighbors of each single country, Fig.~\ref{fig1}B (and Fig.~S3 in the Supplementary Information~\cite{:sm}), which we quantify using the disparity measure $Y(k)$~\cite{Serrano:2007dg,Herfindahl:1959qs,Hirschman:1964ww}. This indicates that, at the same time as countries gained trade partners, they intensified only a few of their trade connections.

These heterogeneities can be exploited to filter out, for each year, a sparse subnetwork representing the relevant structure that remains after eliminating the contingent interactions which overshadow the information contained in the system~\cite{Serrano:2009a,:sm}. These ``backbones'', Fig.~\ref{fig1}A and Table S3, retain most of the countries but the number of trade channels is drastically reduced to the statistically significant ones (see the Supplementary Information, Fig.~S4~\cite{:sm}). At the same time, they preserve pivotal features of the original trade networks: most of the total trade, global connectedness, the small-world property, the heterogeneous degree distributions, and clustering. Interestingly, the correlation between the degrees of countries within the backbones and their GDPs is always extremely high, Fig.~\ref{fig1}C, meaning that these backbones of world trade networks reveal economic size as an underlying variable. 

\begin{figure*}[t]
\centering
\includegraphics[width=16.4cm]{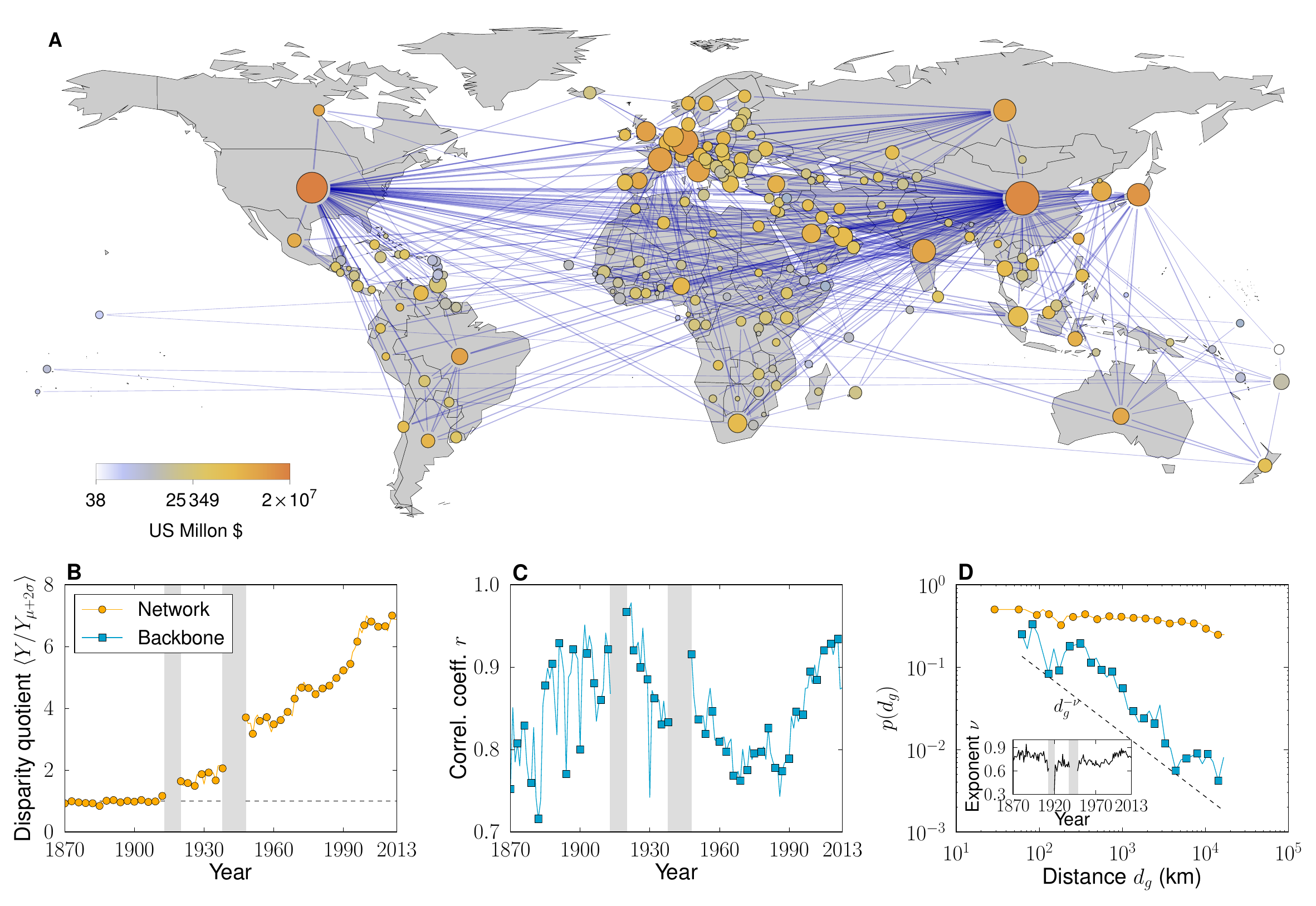} 
\vspace*{.05in}
\caption{{\bf Backbones of International Trade.} {\bf A}: Backbone for the year 2013. Nodes represent countries and their sizes are proportional to the logarithm of the number of trade partners in the backbone. Nodes are colored according to the logarithm of GDP values.  {\bf B}: Evolution of the local heterogeneity level. {\bf C}: Pearson correlation coefficient, $r$, between the GDPs of countries and their degrees in the backbone. In the unfiltered network, values of $r$ are in the range $r\approx 0.3 - 0.4$ after WWII. {\bf D}: Probability of trade as a function of geographic distance for the year 2013. It follows a power law, $p(d) \sim d^{-\nu}$, with $\nu \approx 0.7$ in the backbones; whereas it is almost independent of distance in the unfiltered networks.} 
 \label{fig1}
\end{figure*}
Backbones differ from the unfiltered networks in the relation between geographic distance and trade connections. Overall, the geographic length of connections relative to the geographic distance between all the countries (located at the coordinates of their capitals) shows a sustained increase in the unfiltered network. This means that trade connections become less dependent on distance over time (see the Supplementary Information, Fig.~S2~\cite{:sm}). However, the trade channels within the backbone are comparatively shorter over the whole period, which means that, despite globalization, countries are more likely to choose their more significant trade partners from among their closer neighbors. The connection probability is almost independent of distance in the unfiltered networks, but it shows a clear power-law decay with a stable exponent in the case of the backbones, Fig.~\ref{fig1}D. Strikingly, the filter that produces the backbones is blind to geography. Hence, compared to the unfiltered networks, trade backbones reveal economic size---in terms of GDP---as an underlying variable, and show that geographic distance plays an increasingly important role for the most significant trade channels.

% =================================================================================================
\subsection{WTM construction: a gravity model for international trade channels}
World trade backbones are suitable to be mapped onto an underlying trade space in which closer countries have a greater chance of becoming connected by a trade channel. The likelihood of becoming connected is based on a gravity model consisting of economic size factors and effective trade distance. The simplest metric space that can be considered to represent effective distance is a circle on which countries are separated by an angular distance $d_a=\min(|\Delta \theta|,2\pi-|\Delta \theta|)$  (from now on, effective distance and angular distance are used indistinctly). We propose that the probability that any pair of countries $i$ and $j$ are connected in terms of this distance is
\begin{equation}
p_{ij}=\frac{1}{1+\chi_{ij}^{\beta}},
\label{eq:1}
\end{equation}
where $\chi_{ij}=d_{a,ij}/(\mu\kappa_i \kappa_j)$, such that $p_{ij}$ decreases as a function of $d_{a}$ and increases with the expected economic sizes $\kappa$~\cite{Serrano:2008hb}. These economic sizes are strongly correlated with degrees in the backbone and so with the corresponding GDPs, Fig.~\ref{fig1}A and C. We assume that distance $d_{a}$ incorporates the different factors that shape the complex architecture of the international trade system --not only geography--, and so that are implicitly encoded in its connectivity pattern. The parameters $\mu$ and $\beta$ govern the average degree and clustering, respectively, of the network. Note that $\beta$ is an elasticity measure with respect to trade distances; it calibrates the coupling between the topology and the underlying metric space. Hence, the proposed connection probability resembles Newton's law of gravitation and, therefore, the classical gravity model predicting the volume of bilateral trade flows. Notice, however, that here $p_{ij}$ is not used to predict the volume of trade but the existence of a significant trade channel; information which has to be provided \textit{a priori} in classical gravity models of trade, since they reproduce the existence of a link along with its weight very badly~\cite{Duenas:2013}. 

The gravity model defined by Eq.~\eqref{eq:1} is isomorphic to a purely geometric network model in the hyperbolic plane~\cite{KrPa10}. Hyperbolic geometry has been conjectured as the natural geometry underlying the complex features --{\it e.g.}, scale-free degree distributions, strong clustering levels, the small-world property, {\it etc.}-- of real world networks~\cite{Papadopoulos:2012uq,Krioukov:2009sy}. A possible explanation of such geometric interpretation of complex networks is provided by the gravity law type of connection probability introduced in Eq.~\eqref{eq:1}. Indeed, by mapping the variable $\kappa$ to a radial coordinate $r$ as follows
\begin{equation} \label{mapping}
r=R-2 \ln{\left[ \frac{\kappa}{\kappa_0}\right]}
\; \;\mbox{ with }\; \;
R=2\ln{\left[ \frac{2}{\mu \kappa_0^2}\right]}
\end{equation}
and keeping the same angular coordinates, the connection probability Eq.~\eqref{eq:1} can be rewritten as
\begin{equation}
p_{ij}=\frac{1}{1+ e^{\frac{\beta}{2} (x_{ij}-R)}}
\end{equation}
where $x_{ij}=r_i+r_j+2 \ln{\frac{d_{a,ij}}{2}} \approx d_{h,ij}$ is a very good approximation of the hyperbolic distance $d_{h,ij}$ between two points at radial coordinates $r_i$ and $r_j$ and angular separation $d_{a,ij}$ (the exact hyperbolic distance can be computed with the hyperbolic law of cosines~\cite{:sm}). In Eq.~\eqref{mapping}, $\kappa_0$ is the minimum size of an economy. The hyperbolic representation condensates in a pure geometric framework the properties of the entire system --economic size and effective distance in the gravity model Eq.~\eqref{eq:1}; that is, it allows to draw genuine maps of the trade system where different parts can be compared on an equal footing. More specifically, due to the inverse relation between economic size $\kappa$ and hyperbolic radius $r$, large economies with high degrees are located close to the center of the disk, whereas small economies are placed near its boundary, whereas the distance between nodes at the same angular separation increases with hyperbolic radius.

\begin{figure*}
\centering
\includegraphics[width=15cm]{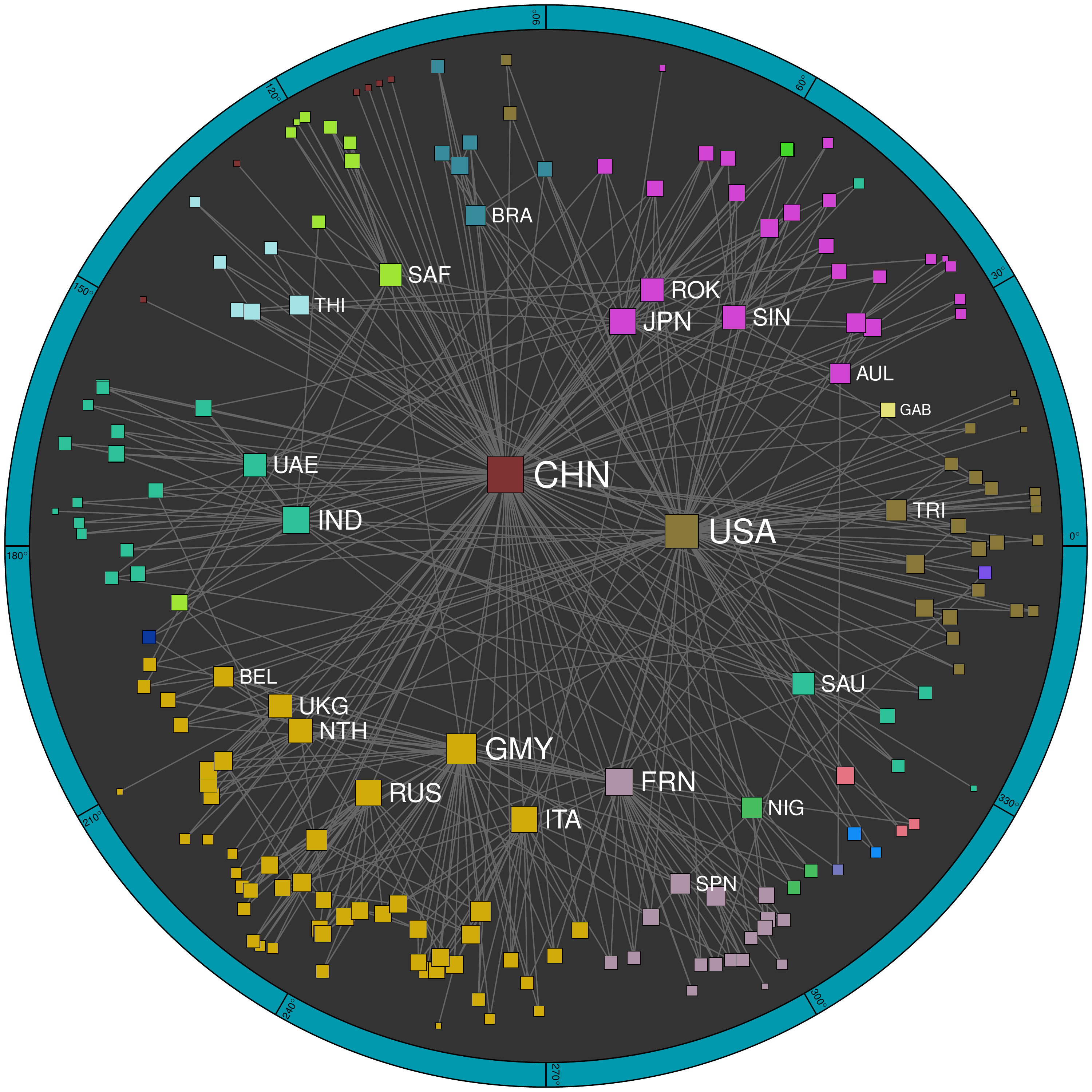} 
\vspace*{.05in}
 \caption{{\bf World Trade Map for 2013.} Representation in the hyperbolic plane. The radial and angular coordinates of countries define the trade distances between them. Symbol sizes are proportional to the logarithm of the degrees of the countries and colors represent communities revealed by the critical gap method presented in Sec.~\ref{communities}. See the Supplementary Information, Table S1~\cite{:sm}, for the country associated with each acronym.}
\label{fig_2}
\end{figure*}

World Trade Maps are constructed by embedding backbones of world trade in hyperbolic space. The embedding method uses statistical inference techniques to identify the coordinates of each country in a backbone which maximize the likelihood that the backbone is reproduced by the model. Very briefly, the likelihood function depends on the connectivity between countries in the backbone and on the probability of connection given by the gravity law in Eq.~\eqref{eq:1}, and so on distances between countries; see the Supplementary Information~\cite{:sm} for more details. Inferred angular distances represent a measure of trade likelihood, except for the (economic) sizes of the countries. This means that two small countries need to be close in terms of angular distance to increase their chance of becoming connected. Inferred angular distances represent a measure of trade likelihood, except for the (economic) 
sizes of the countries. This means that two small countries need to be close in terms of angular distance to increase their chance of becoming connected. The inferred hyperbolic distance, however, incorporates size effects; that is, two countries can be close in hyperbolic space just because of their size. In particular, large countries are in general closer to the rest of the world.

\begin{figure*}[t]
\centering
\includegraphics[width=16.4cm]{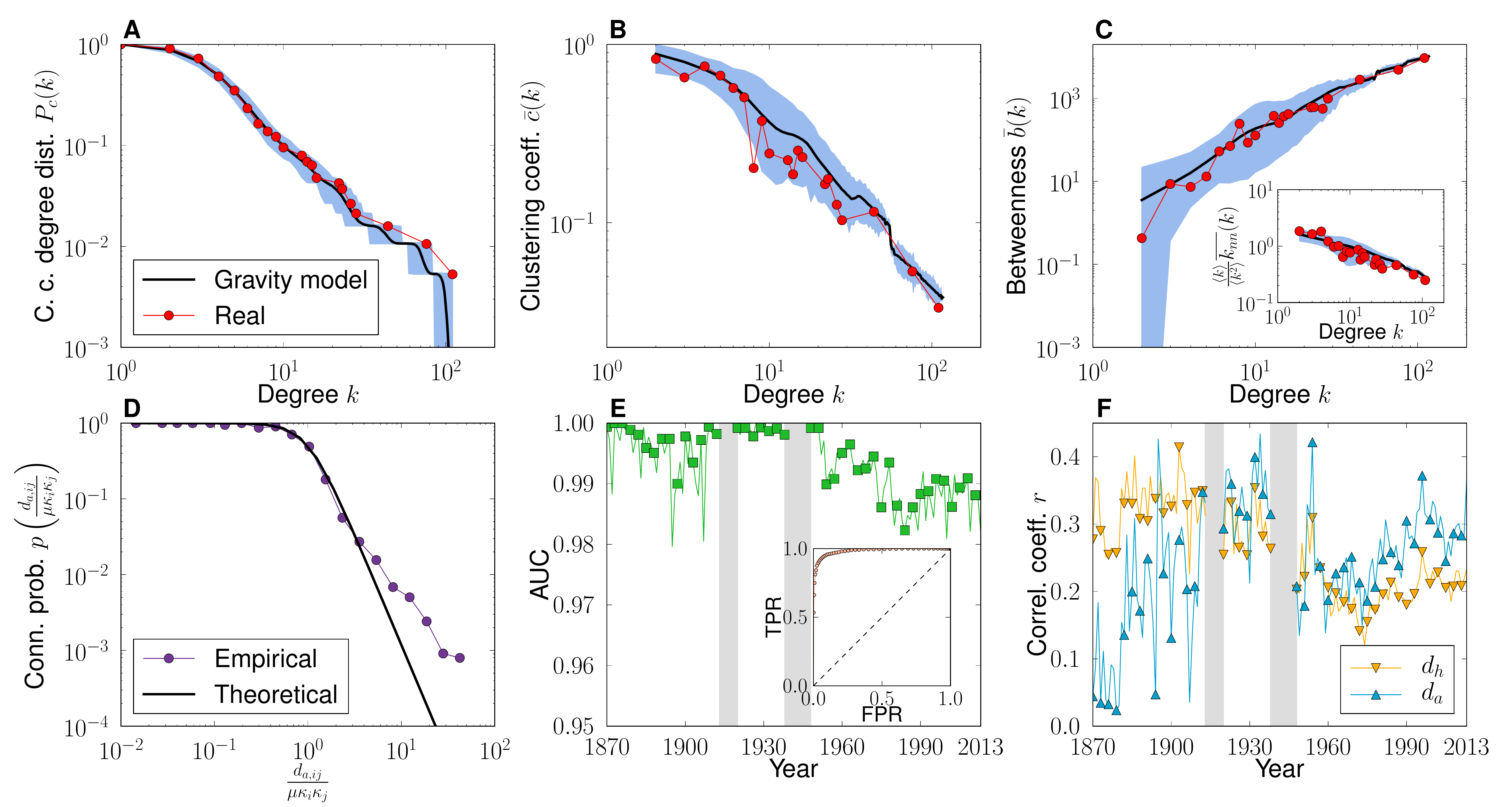} 
\vspace*{.05in}
 \caption{{\bf Quality of the embedding.}  {\bf A-C}: Comparison of the topology of the world trade backbone in 2013 with $1000$ synthetic networks generated by the gravity model of trade channels using the coordinates of the corresponding embedding. Complementary cumulative degree distribution (A), degree-dependent clustering coefficient (B), country betweenness centrality (C), and average nearest-neighbor degree (inset C). The thick black line gives the average over realizations and the error bars mark the $99\%$ confidence interval. {\bf D}: The empirical connection probability as a function of effective distance rescaled by economic sizes, measured using the coordinates of the embedding for the year 2013. The empirical probability is compared to Eq.~\eqref{eq:1}. {\bf E}: The temporal evolution of the area under the ROC curve or AUC statistic. The inset shows the ROC curve of the model in 2013. {\bf F}: The historical evolution of the Pearson correlation coefficient between hyperbolic and angular distances, $d_h$ and $d_a$, and geographical distance $d_g$. }\label{fig_3}
\end{figure*}

In this way, we apply our mapping method to annual networks of world trade from 1870 to 2013 and obtain 129 World Trade Maps that conform the World Trade Atlas 1870-2013. For visualization purposes, we use a single embedding for each year~\cite{:vd}, Fig.~\ref{fig_2} and Table S5; while for the computation of distances, we average over a hundred realizations, Table S6. Fig.~\ref{fig_3}A-C demonstrates the  the power of the gravity model of trade channels to reproduce the topological properties of the real world trade network backbones. This is due, in particular, to the excellent agreement between the empirical and the theoretical probability of connection as a function of the effective distance between countries rescaled by economic size, Fig.~\ref{fig_3}D. As a further demonstration of the quality of the embeddings, Fig.~\ref{fig_3}E displays  the temporal evolution of the area under the receiver operating characteristic (ROC) curve or AUC statistics  measuring the ability of the model in predicting real trade channels and, in the inset, the ROC curve for the year 2013 (see the Supplementary Information, Fig.~S8~\cite{:sm}). Finally, a key observation is that the correlation between embedding distances and geographic distances, as shown in Fig.~\ref{fig_3}F, is significant over the whole period, but far from one; meaning that trade distances encode more than purely geographical information. 

%
%
%
%
%
%
% =================================================================================================
\section*{Long-term evolution of the international trade system}
The World Trade Atlas 1870-2013 is displayed as an interactive tool~\cite{:vd}. It allows us to visualize the evolution of the international trade system over fourteen decades. During the 19th century, the atlas shows a sparsely populated trade space with the hegemonic UK at its core; and it reflects the rise of Germany, France and the USA as they move towards the center prior to WWI. During the interwar period, the atlas presents a central triangle formed by the UK, the USA and Germany; with the USA then progressively becoming the new hegemonic economy during the second part of the 20th century. In the 1960s, decolonization introduces many new countries and at the same time the upper intermediate region becomes more populated by actors such as France, Italy, Japan, and the Netherlands; together with, in less prominent positions, India, Russia, Spain and Belgium, among others. In the 1990s, the European Community starts its process of construction, with Germany, Italy and France coming closer; while the Soviet Bloc remains very close to Russia. During the last few years, China has moved towards a more central position, as a new pretender to superpower status; while the USA, the European Community and Japan have moved to less central positions following the decline of their relative dominance. 
%
%
%

% =================================================================================================
\subsection{Hierarchical organization of the trade system}

\begin{figure*}[t]
\centering
\includegraphics[width=16.4cm]{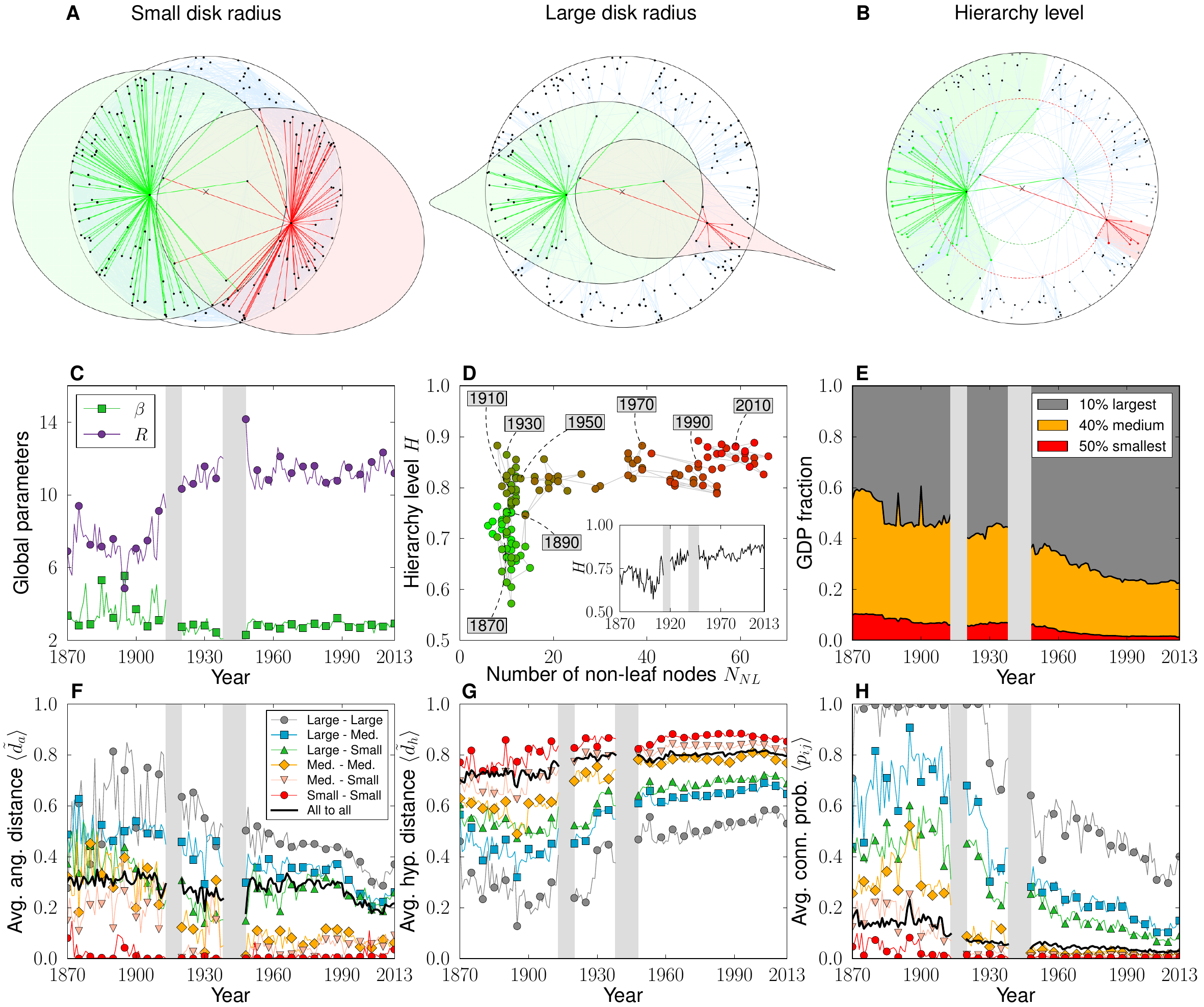} 
\vspace*{.05in}
 \caption{{\bf Hierarchy in World Trade Maps.} {\bf A}: Two hyperbolic disks with different radii (rescaled). Within each disk, we draw two circles of the same radius but centered at different locations. For small radius, the two inner circles appear to be close to Euclidean disks; whereas for large radius, the inner circles are strongly distorted. {\bf B}: Sketch illustrating the definition of the hierarchy level, $H$, of two different nodes. Leaf nodes appear in gray. {\bf C}: Evolution of the radius, $R$, of the hyperbolic disk and of the elasticity parameter, $\beta$. {\bf D}: Evolution of the hierarchy level, $H$, as a function of the number of non-leaf countries and of time (inset). Note that the number of non-leaf countries has grown (with some fluctuations) over time, so that the increase of $N_{NL}$ along the $x$-axis is roughly chronological. {\bf E}: Evolution of the fraction of total GDP accumulated by the top 10\%, 40\%, and 50\% of countries by economic size. {\bf F}--{\bf G}--{\bf H}: Evolution of: the average normalized angular distance (only connected countries), the average normalized hyperbolic distance, and the average connection probability, between countries within the same group and in the different groups defined in {\bf E}. \label{fig_5}}
\end{figure*}

Notice that small economies---low-degree countries---are always located at the periphery of the disk, while large economies---with high degrees---tend to settle at its center. This radial stratification is a sign of the hierarchical organization of the trade system. A rough measure of the hierarchical position of a node is given by its radial coordinate, and so by its degree, as compared to the radius, $R$, of the trade space. This latter measures the hyperbolicity of the whole system or, equivalently, its departure from flatness, Fig.~\ref{fig_5}A. $R$ grows until reaching a stationary value at the end of the interwar period, around which it fluctuates thereafter. Interestingly, the opposite behavior is observed for the elasticity parameter, $\beta$, Fig.~\ref{fig_5}C and Table S4. Larger values of $R$ imply that differences in trade distances are more important for peripheral nodes; that is, they connect to each other less frequently.  

To study this issue further, we define the level of hierarchy as a scalar, $H$, based on angular distances. For each non-leaf country $i$ with radius $r_i$---a non-leaf country has at least one trading partner $j$ whose economy is smaller, i.e, with a radial position such that $r_j > r_i$---we measure the average angular separation $\overline{d_{a}}_{,i}$ with its trading partners for which $r>r_i$, and define its hierarchical level as $h_i= 1-2 \overline{d_{a}}_{,i}/ \pi$, Fig.~\ref{fig_5}B. If the neighbors have exactly the same angular coordinates as $i$, $h_i=1$; whereas $h_i=0$ for random angular positions. The global hierarchy level, $H$, is obtained by averaging $h_i$ over all non-leaf countries. In the long run, $H$ increases from below 0.7 in the 19th century to very close to 0.9 in 2013, Fig.~\ref{fig_5}D inset. This represents a substantial increase and situates the system at extremely high levels of hierarchy. The evolution of $H$ as a function of the number of non-leaf countries, Fig.~\ref{fig_5}D, is even more revealing. Even if the number of non-leaf countries has increased noticeably in the last decades, $H$ has not decreased but has increased. This indicates an expansion of the depth of the hierarchy, from a quasi star-like organization before WWI to a deeper hierarchical structure after WWII, with more countries at intermediate layers acting as local economic hubs.

The economic significance of this effect can be explored further by analyzing the evolution of trade distances between countries classified according to their economic size. We rank countries in decreasing order of GDP and divide them into three groups: the top $10\%$, the middle $40\%$, and the bottom $50\%$, Fig.~\ref{fig_5}E. Fig.~\ref{fig_5}F, G and H shows the evolution of: the average angular distance (between connected pairs); the average hyperbolic distance; and the average connection probability, for all pairs of countries within the same group and in the different groups. Hyperbolic and angular distances are normalized by the diameter of the space, $2R$ and $\pi$, respectively, so that different years are comparable. There is a sharp and persistent stratification according to economic group, denoting a positive correlation between trade distance, or connection probability, and economic status. The average angular distance strongly correlates with economic size. Interestingly, we observe the opposite behavior for the average hyperbolic distance: it is greater, the lower the economic size. This reversal reflects the different interpretations of the two trade distance measures: the latter governs the formation of trade channels, and the former is an indicator of clusterization. The hyperbolic distance between economic groups has been steadily increasing since the 1960's; and this increase has been more pronounced for the large economies. In contrast, the angular distance between groups, fluctuating until the end of WWII, has remained very low or decreased over time. This denotes a trend towards concentration of countries in specific regions of the trade space. 
 
Placed in historical perspective, these results indicate that small countries at the bottom of the hierarchy are far from the rest of the world in terms of trade distance. They therefore encounter greater difficulties establishing trade channels with other countries, except for the largest economies at the top of the hierarchy, which have more chances of becoming connected worldwide. At the same time, the increase in hyperbolic distance may be a consequence of the increase in hierarchy and of market competition effects; while the decrease in angular distance indicates the formation of better-defined trading blocs that form communities in the trade space. 
%
%
%
% =================================================================================================
\subsection{Natural communities based on trade distances\label{communities}}
\begin{figure*}[t]
\centering
\includegraphics[width=16.4cm]{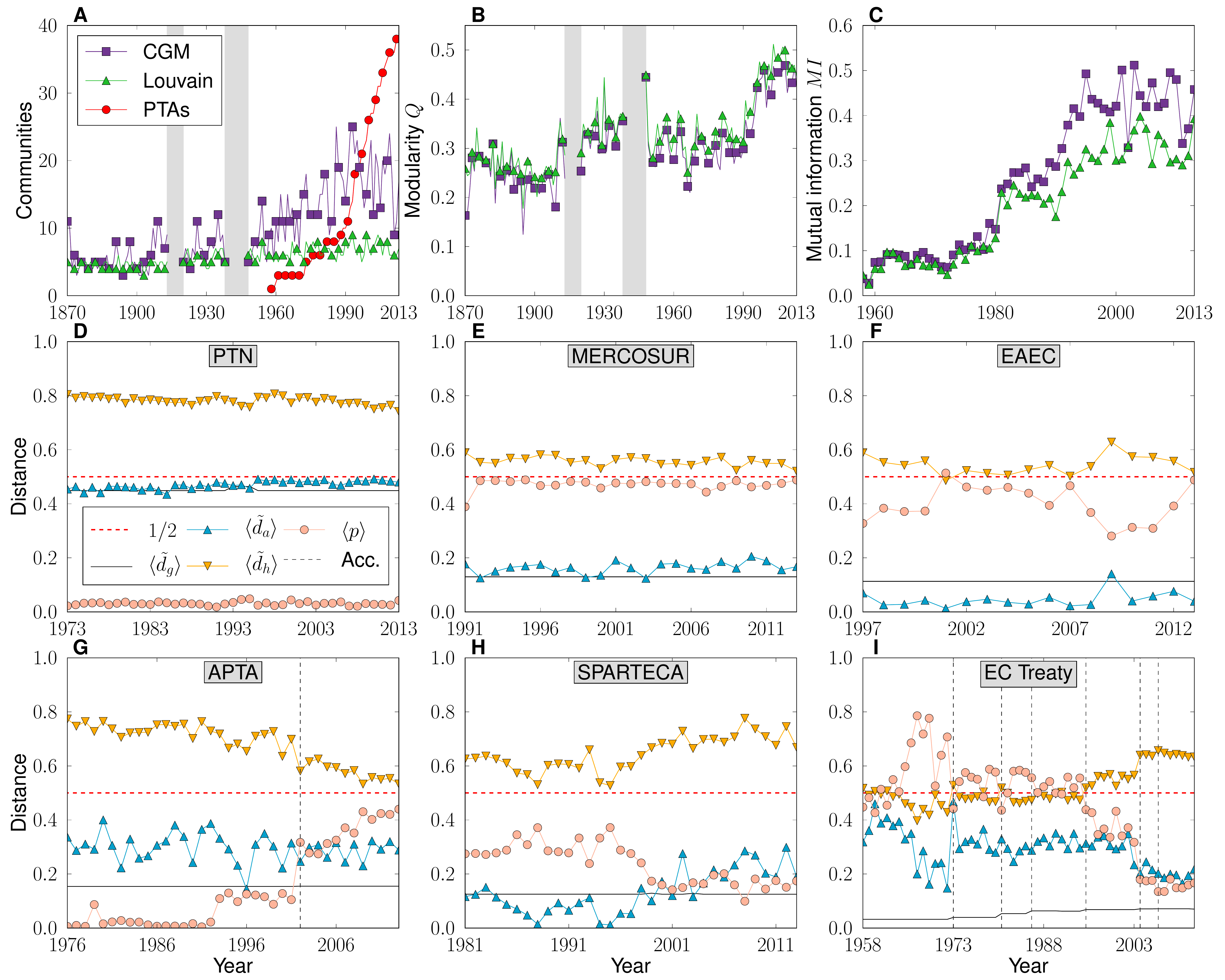} 
\vspace*{.05in}
 \caption{{\bf Natural Communities vs. Preferential Trade Agreements.} {\bf A}: Evolution of the number of communities found by the Louvain method and the critical gap method, along with the evolution of the number of PTAs. {\bf B}: Evolution of the modularity, $Q$, for the two methods. {\bf C}: Mutual information between critical gap method or Louvain communities and PTAs. {\bf D}--{\bf I}: Evolution of normalized angular and hyperbolic distances between countries in representative PTAs, along with the average geographic distance and the average connection probability within each PTA. Dashed vertical lines indicate the accession of new countries. The dashed horizontal red line indicates the $0.5$ level, which, in the case of angular distances, corresponds to a random distribution of points. \label{fig_6}}
\end{figure*}

A precise definition of communities in trade space can be given to subsets of countries that form densely populated zones separated by void angular sectors~\cite{:sm}. We use a critical gap method~\cite{Serrano:2012we} (CGM) to search for communities in the WTMs. The method works by grouping countries in the same community if the angular separation between pairs is smaller than a critical angular distance, which fixes a unique partition into non-overlapping communities. We select the critical gap that yields the maximum congruency with topological communities in the backbone (as measured by modularity~\cite{Newman:2004aa}, $Q$, giving the quality of the division into clusters). Fig.~\ref{fig_2} shows the CGM communities in the WTM for the year 2013 in different colors; see also Table S7. The evolution of the number of CGM communities is shown in Fig.~\ref{fig_6}A, as compared to those obtained by applying the Louvain algorithm. In terms of modularity, Fig.~\ref{fig_6}B, the two methods coincide to an extremely high degree. However, the number of communities discovered by CGM is typically twice that of the Louvain method, since Louvain modules may integrate smaller CGM communities. Both modularity and the number of communities increase over time, reaching a maximum at the beginning of the last economic crisis around 2007, with a minor downturn afterwards. 

This tendency towards localism in trade space seems consistent with the proliferation, since the late 1980's, of PTAs as formal trading blocs~\cite{Krugman:1993,Frankel:1997}, Fig.~\ref{fig_6}A. We use the WTO's Regional Trade Agreements Information System (RTA-IS)~\cite{WTO} to list all plurilateral PTAs in force in 2013 and compare them with natural CGM trade communities. To measure their similarity, we used normalized mutual information~\cite{Cover2006} (MI), which takes a value of $1$ when the two compared partitions are identical and $0$ when there is no more than random matching. Over the entire period, natural communities are noticeably more congruent with PTAs than those identified by the Louvain method, Fig.~\ref{fig_6}C. However, the overlap between PTAs and natural communities is not perfect. 

This poses a question regarding the progression of barriers to trade within each PTA, which can be assessed from the evolution of the average normalized angular and hyperbolic distances---$\langle \tilde{d}_a \rangle$ and $\langle \tilde{d}_h \rangle$, respectively---between its members. We focus in PTAs with at least ten years of history---$31$ trading blocs in 2013---for which we find different patterns of evolution; see representative examples in Fig.~\ref{fig_6}D-I (see also Supplementary Information, Fig.~S12~\cite{:sm}). Both average normalized angular and hyperbolic distances remain stable over time in about half ($17$) of the PTAs, with $\langle \tilde{d}_a \rangle$ typically close but above the geographic value. In some blocs, $\langle \tilde{d}_a \rangle$ and $\langle \tilde{d}_g \rangle$ are extremely congruent, specifically in PTAs with a worldwide composition, Fig.~\ref{fig_6}D; or, at the other extreme, with a strong geographical orientation, Fig.~\ref{fig_6}E. Strikingly, the value of $\langle \tilde{d}_a\rangle$ is found to be below the geographical average in PTAs interconnecting Russia and the republics of the former Soviet Union, which denotes communities with reduced trade barriers, even below levels expected given their geographical dispersion, Fig.~\ref{fig_6}F. 

Distances in the rest of the PTAs show mixed progressions. An interesting case is the Asia-Pacific Trade Agreement (APTA), Fig.~\ref{fig_6}G. For the APTA, $\langle \tilde{d}_h \rangle$ has been steadily decreasing since its inception, even after the accession of China in 2002; whereas $\langle \tilde{d}_a \rangle$ has remained fairly stable. This implies a significant increase in the internal connectivity of the PTA, which may be related to the increasing economic size of their members. In other interesting cases, both  distances present a slight but clear increase; denoting a trend towards trade diversion among their members, as in the South Pacific Regional Trade and Economic Cooperation Agreement (SPARTECA), Fig.~\ref{fig_6}H, mainly formed of small islands in the Pacific Ocean. 

The European Community (EC) Treaty is the largest and most complex of all the PTAs, accounting for $23.6 \%$ of world GDP in 2013. Its evolution has moved through three different stages, Fig.~\ref{fig_6}I. The first lasted from its inception until 1972, just before the accession of Denmark, Ireland, and the UK. During this period, both angular and hyperbolic distances reduced significantly and the internal connectivity increased, which indeed indicates an important reduction of barriers to trade. The aforementioned accessions in 1973 reversed the trend, which indicates that the new members were not natural partners of the treaty before they entered the agreement. During the second stage, 1973--2003, many medium/large economies joined, but internal distances remained stable since these new countries were, {\it de facto}, natural trade partners of the treaty. Between 2003 and 2004, another transition took place with the accession of a large number of small economies, mainly in Eastern Europe. During the transition, $\langle \tilde{d}_a \rangle$ decreased significantly at the same time as $\langle \tilde{d}_h\rangle$ increased, and so the internal connectivity decreased. Before joining the EC these newly adjoined countries already had members of the EC as their main trading partners. After joining the EC, these new countries kept the same main trading partners but, due to their small size, did not increase trade with them. So, the effect has been augmented internal barriers to trade in the last decade, as the increased localization due to the addition of small economies could not be compensated for due to the heterogeneity in economic sizes of the members. 
%
%
%
%
%
% =================================================================================================
\section*{Discussion}
The  gravity law not only models the volume of flows in bilateral trade, but also, as we have proved, the large-scale architecture of the connections within the international trade system.  The gravity model for trade channels works in a trade space in which distance is an effective  aggregated measure which brings together and integrates the different dimensions that shape the international trade network, implicitly encoded in the backbone of its complex connectivity structure. We have proved that the natural  geometric representation that combines into a single trade distance the effects of economic size and effective distance is hyperbolic space, which we used to construct the World Trade Atlas 1870-2013  summarising international trade history between these dates. Most importantly, as stated before by others in an interesting discussion of globalization~\cite{friedman_book,stiglitz_book,Pankaj_2007}: the world is not flat. We claim, on scientific grounds, that it is hyperbolic, as hyperbolic space is the natural geometry having the power to embed its complex structural organization. 

In contrast to the widespread perception that globalization has led to a decrease in the importance of distance, we observe that countries preferentially select their significant trade partners from geographically close neighbors, in line with general statistics~\cite{Ghemawat_2013}. According to the World Trade Atlas 1870-2013, the role of trade distance has not decreased but increased over time, driven by two main forces. First, we report an increased hierarchical organization related to a persistent stratification by economic size, so that not all global trade market competitors have equal opportunities. It was reported previously that elasticity with respect to distance of bilateral trade between high-income countries fell in the period 1962-1996~\cite{Brun:2005}, with no trend for the group of low-income countries. The portrait that emerges from our maps is more dire. Differences in trade distances are becoming more important, particularly for small economies at the bottom of the hierarchy, which are moving away from the rest of the world. The small economies are encountering more difficulties in establishing trade channels, except for those with the largest economies at the top of the hierarchy, which have more chance of becoming connected worldwide. 

Second, we observe a movement towards localism---already encountered before in the context of economic geography~\cite{krugman:1991gc,fujita1999}---as a tendency to concentrate within natural trade communities. Interestingly, this trend seems to have been reverted since 2009, maybe as a consequence of the fast rise of China as a new commercial power and due to the effects of the economic crisis. Despite the proliferation of PTAs as formal trading blocs, we only found a moderate overlap with natural communities. Indeed, PTAs have not necessarily reduced barriers to trade between their members, as measured by hyperbolic and angular distances. These results reveal PTAs as a tool that may serve purposes other than trade in economics or politics, so that their ambiguous consequences on the creation or steering of trade depend upon several other conditions~\cite{Krugman:1993,Rosson_1994}. In our framework, we observe that the localization effect is entangled with that of hierarchization; that is, with the formation of intermediate hubs that dominate well-delimited angular regions as the number of layers in the hierarchy grows. Both effects, as two sides of the same coin, may have been exacerbated by trade liberalization policies with uneven effects among non-equals~\cite{cage_2012}.

Our discussion has focused on revealing, through a single historical picture, globalization, hierarchization, and localization as the main forces shaping the trade space, which far from being flat is hyperbolic. The World Trade Atlas 1870-2013 can help to shed light on a number of other questions based on trade distances, for instance, those regarding the specific composition of natural communities in trade space, or it can facilitate a new approach to the analysis of the relationship between trade and other economic factors. 

\begin{acknowledgments}
This work was supported by: a James S. McDonnell Foundation Scholar Award in Complex Systems; the ICREA Academia award, funded by the {\it Generalitat de Catalunya}; the MINECO project no.\ FIS2013-47282-C2-1-P; the {\it Generalitat de Catalunya} grant no.\ 2014SGR608; and the European Commission FET-Proactive Project MULTIPLEX (grant 317532). 
\end{acknowledgments}

\end{document}